\documentclass[12pt,a4paper]{article}
\usepackage{fullpage}
\usepackage{cite}
\usepackage[centertags]{amsmath}
\allowdisplaybreaks[4]
\numberwithin{equation}{section}
\usepackage{amsbsy}
\usepackage{amsfonts}
\usepackage{amssymb}
\usepackage{url}
\usepackage[dvips]{hyperref}
\newcommand{\vet}[1]{\ensuremath{\hskip-1pt\vec{\hskip1pt#1}}}

\begin{document}

\begin{flushright}
\textsf{30 January 2004}
\\
\textsf{hep-ph/0401244}
\end{flushright}

\vspace{1cm}

\begin{center}
\large
\textbf{Theory of Neutrino Oscillations}\footnote{Talk presented at the
11th Lomonosov Conference on Elementary Particle Physics,
21--27 August 2003, Moscow State University, Moscow, Russia.}
\normalsize
\\[0.5cm]
\large
Carlo Giunti
\normalsize
\\[0.5cm]
INFN, Sezione di Torino, and Dipartimento di Fisica Teorica,
\\
Universit\`a di Torino,
Via P. Giuria 1, I--10125 Torino, Italy
\\[0.5cm]
\begin{minipage}[t]{0.8\textwidth}
\begin{center}
\textbf{Abstract}
\end{center}
We review critically the main assumptions on which the standard theory
of neutrino oscillations is based.
We show that all assumptions are realistic, except the so-called
``equal momentum assumption'',
which however is irrelevant.
We briefly review the covariant plane-wave derivation of neutrino oscillations
and a quantum field theoretical wave packet model of neutrino oscillations.
We show that both approaches
lead to the standard expression for the oscillation phase.
The wave packet model allows also
to describe the coherence of the oscillations
and the localization of the production
and detection processes.
\end{minipage}
\end{center}

\section{Introduction}
\label{Introduction}

The possibility of
neutrino oscillations was discovered
by Bruno Pontecorvo in the late 50's
following an analogy with kaon oscillations
\cite{Pontecorvo:1957cp,Pontecorvo:1958qd}.
Since at that time only one \emph{active} neutrino was known,
Pontecorvo invented the concept of a \emph{sterile} neutrino \cite{Pontecorvo:1968fh},
which is a neutral fermion which does not take part to
weak interactions.
The muon neutrino was discovered
in 1962
in the Brookhaven experiment
of Lederman, Schwartz, Steinberger \textit{et al.}
\cite{Danby:1962nd},
which followed a proposal made by Pontecorvo in 1959
\cite{Pontecorvo-neutrinos-59}.
Then,
it became clear that oscillations
between different active neutrino flavors are possible
if neutrinos are massive and mixed particles.
Indeed,
in 1967 \cite{Pontecorvo:1968fh}
Pontecorvo predicted the solar neutrino problem
as a possible result of $\nu_e \to \nu_\mu$
(or $\nu_e \to \nu_{\text{sterile}}$)
transitions
before the first measurement of the
solar electron neutrino flux in the
Homestake experiment
\cite{Cleveland:1998nv},
and in 1969 Gribov and Pontecorvo
discussed in detail
the possibility of solar neutrino oscillations
due to neutrino mixing
\cite{Gribov:1969kq}.

However,
in these and other papers written before 1976
the probability of neutrino oscillations
was not calculated in a rigorous way,
but simply estimated
on the basis of the analogy with kaon oscillations.
As a result,
the phase of the oscillations
was correct within a factor of two.

The standard theory of neutrino oscillations
was developed in 1976 by
Eliezer and Swift
\cite{Eliezer:1976ja},
Fritzsch and Minkowski
\cite{Fritzsch:1976rz},
Bilenky and Pontecorvo
\cite{Bilenky:1976yj}
(see the beautiful review in Ref.~\cite{Bilenky:1978nj})
on the basis of the
following four assumptions that will be discussed critically in
this report:

\renewcommand{\labelenumi}{\theenumi}
\renewcommand{\theenumi}{(A\arabic{enumi})}
\begin{enumerate}
\item
\label{A1}
Neutrinos are ultrarelativistic particles.
\item
\label{A2}
Neutrinos produced or detected in
CC weak interaction processes
are described by the
flavor states
\begin{equation}
|\nu_{\alpha}\rangle
=
\sum_{k}
U_{{\alpha}k}^{*}
\,
|\nu_{k}\rangle
\,,
\label{001}
\end{equation}
where
$U$ is the unitary mixing matrix,
$\alpha=e,\mu,\tau$,
and
$|\nu_{k}\rangle$ is the state of a neutrino with mass $m_k$.
\item
\label{A3}
The propagation time
is equal to the
distance $L$
traveled by the neutrino
between production and detection.
\item
\label{A4}
The massive neutrino states
$|\nu_{k}\rangle$
in Eq.~(\ref{001})
have the same momentum,
$p_k = p \simeq E$
(``equal momentum assumption''),
and different energies,
$
E_k
=
\sqrt{ p^2 + m_k^2 }
\simeq
E
+
m_k^2/2E
$,
where $E$ is the neutrino energy neglecting mass effects
and the approximations are valid for ultrarelativistic neutrinos.
\end{enumerate}

In Section~\ref{Standard theory of neutrino oscillations}
we briefly review the main points of the
standard theory of neutrino oscillations.
In Sections~\ref{ultrarelativistic neutrinos}--\ref{equal momentum}
we discuss critically the four assumptions listed above.
In Section~\ref{Covariant derivation of neutrino oscillations}
we review the
covariant derivation of the neutrino oscillation probability
in the plane wave approach
\cite{Winter:1981kj,Giunti:2000kw,Bilenkii:2001yh,hep-ph/0311241}.
In Section~\ref{Wave packet model}
we review the quantum field theoretical
wave packet model
presented in Ref.~\cite{Giunti:2002xg}.
Finally,
in Section~\ref{Conclusions}
we present our conclusions.

\section{Standard theory of neutrino oscillations}
\label{Standard theory of neutrino oscillations}

In the plane wave approximation
the states $|\nu_k\rangle$ of massive neutrinos
are eigenstates of the free Hamiltonian
with definite energy eigenvalues $E_k$.
Therefore,
their time evolution is given by the Schr\"odinger equation,
whose solution is
\begin{equation}
|\nu_{k}(t)\rangle
=
e^{-iE_kt}
\,
|\nu_{k}\rangle
\,.
\label{101}
\end{equation}
Using assumption~\ref{A2},
from Eq.~(\ref{001})
the time evolution of the flavor states is given by
\begin{equation}
|\nu_{\alpha}(t)\rangle
=
\sum_{\beta=e,\mu,\tau}
\left(
\sum_{k}
U_{{\alpha}k}^{*}
\,
e^{-iE_kt}
\,
U_{{\beta}k}
\right)
|\nu_{\beta}\rangle
\,,
\label{102}
\end{equation}
which,
for $t>0$,
is a superposition of different flavors
if the mixing matrix is non-diagonal.
The coefficient of the flavor state $|\nu_{\beta}\rangle$
is the amplitude of
$\nu_\alpha\to\nu_\beta$
transitions,
whose squared absolute value gives the probability
\begin{equation}
P_{\nu_{\alpha}\to\nu_{\beta}}(t)
=
\left|
\langle\nu_{\beta}|\nu_{\alpha}(t)\rangle
\right|^2
=
\left|
\sum_{k}
U_{{\alpha}k}^{*}
\,
e^{-iE_kt}
\,
U_{{\beta}k}
\right|^2
\,.
\label{103}
\end{equation}
Using the equal-momentum assumption~\ref{A4},
the energy of the $k^{\text{th}}$
massive neutrino component
is given by
$E_k=\sqrt{p^2+m_k^2}$,
which can be approximated to
$E_k \simeq p + m_k^2/2p$
in the case of ultrarelativistic neutrinos
following from assumption~\ref{A1}.
Moreover,
assumption~\ref{A3}
allows to replace the usually unknown propagation time $t$ with
the usually known distance $L$
traveled by the neutrino
between production and detection.
The final result for the oscillation probability
can be written as
\begin{equation}
P_{\nu_{\alpha}\to\nu_{\beta}}(L)
=
\sum_{k}
|U_{{\alpha}k}|^2
|U_{{\beta}k}|^2
+
2
\mathrm{Re}
\sum_{k>j}
U_{{\alpha}k}^{*}
U_{{\beta}k}
U_{{\alpha}j}
U_{{\beta}j}^{*}
\exp\!\left(
-i\frac{\Delta{m}^2_{kj} L}{2E}
\right)
\,,
\label{104}
\end{equation}
where
$ \Delta{m}^2_{kj} = m_k^2 - m_j^2 $
and
$E=p$ is the neutrino energy
neglecting mass contributions.
In Eq.~(\ref{104})
we have separated the expression for the flavor transition probability 
into a constant term
and a term which oscillates as a function of the distance $L$.
The oscillating term is the most interesting one from a
quantum mechanical point of view,
because it is due to the interference between
the different massive neutrino components,
whose existence requires coherent
production and detection.
On the other hand,
the constant term
is experimentally very important,
because it gives the average probability
of flavor transitions,
which is the measured one when
the oscillating term is not present because of lack of coherence
or when
the oscillating term is not measurable because it is washed out
by the average over the energy resolution of the detector
or the distance uncertainty.

Let us now examine critically one by one
the four assumptions \ref{A1}--\ref{A4}
that leaded to the result (\ref{104}).

\section{Assumption \ref{A1}: ultrarelativistic neutrinos}
\label{ultrarelativistic neutrinos}

The assumption \ref{A1} is correct, because neutrino masses
are smaller than about one eV
(see Refs.~\cite{Bilenky:2002aw,hep-ph/0310238})
and only neutrinos with energy larger than about 100 keV
can be detected.

Indeed,
neutrinos are detected in:
\renewcommand{\labelenumi}{\theenumi.}
\renewcommand{\theenumi}{\arabic{enumi}}
\begin{enumerate}
\item
Charged-current or neutral-current weak processes
which have an energy threshold
larger than some fraction of MeV.
This is due to the fact that
in a scattering process
$
\nu + A \to \sum_X X
$
with $A$ at rest,
the squared center-of-mass energy
$s = 2 E_\nu m_A + m_A^2$
(neglecting the neutrino mass)
must be bigger than
$( \sum_X m_X )^2$,
leading to
$
E_{\nu}^{\mathrm{th}}
=
\frac{ ( \sum_X m_X )^2 }{ 2 m_A } - \frac{ m_A }{ 2 }
$.
For example:
\renewcommand{\labelitemi}{$-$}
\begin{itemize}
\item
$
E_{\nu}^{\mathrm{th}}
\simeq
0.233 \, \mathrm{MeV}
$
for
$ \nu_e + {}^{71}\mathrm{Ga} \to {}^{71}\mathrm{Ge} + e^- $
in gallium
solar neutrino experiments
(see Ref.~\cite{hep-ph/0310238}).
\item
$
E_{\nu}^{\mathrm{th}}
\simeq
0.81 \, \mathrm{MeV}
$
for
$ \nu_e + {}^{37}\mathrm{Cl} \to {}^{37}\mathrm{Ar} + e^- $
in the Homestake \cite{Cleveland:1998nv}
solar neutrino experiment.
\item
$
E_{\nu}^{\mathrm{th}}
\simeq
1.8 \, \mathrm{MeV}
$
for
$ \bar\nu_e + p \to n + e^+ $
in reactor neutrino experiments
(see Ref.~\cite{hep-ph/0310238}).
\item
$
E_{\nu}^{\mathrm{th}}
\simeq
2.2 \, \mathrm{MeV}
$
in the neutral-current process
$ \nu + d \to p + n + \nu $
used in the SNO experiment to detect active solar neutrinos
\cite{Ahmad:2002jz}.
%
%
\end{itemize}
\item
The elastic scattering process
$ \nu + e^- \to \nu + e^- $,
whose cross section is proportional to the neutrino energy
($
\sigma(E_{\nu})
\sim
\sigma_0 E_{\nu} / m_e
$,
with
$
\sigma_0
\sim
10^{-44} \, \mathrm{cm}^2
$).
An energy threshold
of some MeV's
is needed in order to have a signal above the background.
For example,
$
E_{\nu}^{\mathrm{th}}
\simeq
5 \, \mathrm{MeV}
$
in the Super-Kamiokande \cite{SK-sun-01}
solar neutrino experiment.
\end{enumerate}

As we will see, the ultrarelativistic character of neutrinos
implies the correctness of the assumptions
\ref{A2} and \ref{A3}
and the irrelevance of the assumption
\ref{A4},
which is not realistic.

\section{Assumption \ref{A2}: flavor states}
\label{flavor states}

In Ref.~\cite{Giunti:1992cb} it has been shown that
the assumption \ref{A2} is not exact,
because the amplitude of production and detection
of the massive neutrino $\nu_k$
is not simply given by $U_{{\alpha}k}^{*}$
(see also Refs.~\cite{Bilenkii:2001yh,Giunti:2002xg}).
However,
in the ultrarelativistic approximation
the characteristics of the production and detection processes
that depend on the neutrino mass can be neglected,
leading to a correct approximate description of flavor neutrinos
through the states (\ref{001}).

\section{Assumption \ref{A3}: $t=L$}
\label{tL}

The assumption \ref{A3} follows from the ultrarelativistic approximation,
because ultrarelativistic particles propagate almost at the velocity of light.
However,
in the standard theory of neutrino oscillations
massive neutrinos are treated as plane waves,
which are limitless in space and time.
In order to justify the assumption \ref{A3}
it is necessary to treat massive neutrinos as wave packets
\cite{Giunti-Kim-Lee-Whendo-91},
which are localized on the production process
at the production time and
propagate between the production and detection processes at a velocity
close to the velocity of light.
Such a wave packet treatment
\cite{Giunti-Kim-Lee-Whendo-91,Giunti-Kim-Coherence-98,Giunti:2002xg,Giunti:2003ax}
yields the standard formula for the oscillation length.
In addition,
the different group velocities of different massive neutrinos
imply the existence of a coherence length for the oscillations,
beyond which the wave packets of different massive neutrinos
do not jointly overlap with the detection process
\cite{Nussinov:1976uw,Kiers:1996zj}.

The wave packet treatment of neutrino oscillations
is also necessary for a correct description
of the momentum and energy uncertainties
necessary for the coherent production and detection of
different massive neutrinos
\cite{Kayser:1981ye,Beuthe:2001rc,Giunti:2003ax},
whose interference generates the oscillations.

The physical reason why the substitution $t = L$ is correct
can be understood by noting that,
if the massive neutrinos are ultrarelativistic
and contribute coherently to the detection process,
their wave packets
overlap with the detection process
for an interval of time $[ t - \Delta t \,,\, t + \Delta t ]$,
with
\begin{equation}
t
=
\frac{L}{\overline{v}}
\simeq
L \left( 1 + \frac{\overline{m^2}}{2E^2} \right)
\,,
\qquad
\Delta t \sim \sigma_x
\,,
\label{505}
\end{equation}
where $\overline{v}$
is the average group velocity,
$\overline{m^2}$ is the average of the squared neutrino masses,
$\sigma_x$
is given by the spatial uncertainties of the production and detection processes
summed in quadrature
\cite{Giunti-Kim-Coherence-98}
(the spatial uncertainty of the production process
determines the size of the massive neutrino wave packets).
The correction $ L \overline{m^2} / 2E^2 $ to $t=L$
in Eq.~(\ref{505})
can be neglected,
because it gives corrections to the oscillation phases
which are of higher order in the very small ratios
$ m_k^2 / E^2 $.
The corrections due to
$\Delta t \sim \sigma_x$
are also negligible,
because in all realistic experiments
$ \sigma_x $
is much smaller than the oscillation length
$L^{\mathrm{osc}}_{kj} = 4 \pi E / \Delta{m}^2_{kj}$,
otherwise oscillations could not be observed
\cite{Kayser:1981ye,Giunti-Kim-Lee-Whendo-91,Beuthe:2001rc,Giunti:2003ax}.
One can summarize these arguments by saying that
the substitution $t = L$ is correct
because
the phase of the oscillations
is practically constant over the interval of time in which the
massive neutrino wave packets overlap with the detection process
and it is given by
\begin{equation}
\phi_{kj}(L)
=
\frac{ \Delta{m}^2_{kj} L }{ 2 E }
=
2 \pi \,
\frac{L}{L^{\mathrm{osc}}_{kj}}
\,,
\label{503}
\end{equation}
plus negligible corrections of higher order in the neutrino masses.

\section{Assumption \ref{A4}: equal momentum}
\label{equal momentum}

Let us discuss now the assumption \ref{A4},
which has been shown to be unrealistic in
Refs.~\cite{Giunti:2001kj,Giunti:2003ax}
on the basis of simple relativistic arguments.
Indeed, the relativistic transformation of energy and momentum
implies that the equal momentum assumption
cannot hold concurrently in different inertial systems.
On the other hand,
the probability of flavor neutrino oscillations is independent from the
inertial system adopted for its measurement,
because the neutrino flavor is measured by observing charged leptons whose character
is Lorentz invariant
(\textit{e.g.} an electron is seen as an electron in any system of reference).
Therefore,
the probability of neutrino oscillations
is Lorentz invariant
\cite{Giunti:2000kw,physics/0305122}
and must be derived in a covariant way.
In fact,
the oscillation probability has been derived
without special assumptions about the energies and momenta of
the different massive neutrino components
both in the plane wave approach
\cite{Winter:1981kj,Giunti:2000kw,Bilenkii:2001yh,hep-ph/0311241}
and in the wave packet treatment
\cite{Giunti-Kim-Lee-Whendo-91,Giunti-Kim-Coherence-98,Giunti:2001kj,Giunti:2003ax}.

\section{Covariant derivation of neutrino oscillations}
\label{Covariant derivation of neutrino oscillations}

Let us briefly describe the covariant derivation of the neutrino oscillation probability
in the plane wave approach, in which
the massive neutrino states in Eq.~(\ref{001})
evolve in space and time as plane waves:
\begin{equation}
|\nu_{k}(x,t)\rangle
=
e^{- i E_k t + i p_k x}
\,
|\nu_{k}\rangle
\,.
\label{002}
\end{equation}
Substituting Eq.~(\ref{002}) in Eq.~(\ref{001})
and expressing the $|\nu_{k}\rangle$ on the right-hand side in terms of flavor states
($
|\nu_{k}\rangle
=
\sum_{\beta=e,\mu,\tau}
U_{{\beta}k}^*
\,
|\nu_{\beta}\rangle
$),
we obtain
\begin{equation}
|\nu_{\alpha}(x,t)\rangle
=
\sum_{\beta=e,\mu,\tau}
\left(
\sum_{k}
U_{{\alpha}k}
\,
e^{- i E_k t + i p_k x}
\,
U_{{\beta}k}^*
\right)
|\nu_{\beta}\rangle
\,,
\label{003}
\end{equation}
which shows that
at a distance $x$ and after a time $t$
from the production of a neutrino with flavor $\alpha$,
the neutrino is a superposition of
different flavors
(if the mixing matrix is not diagonal).
The probability of flavor transitions in space and time is given by
\begin{equation}
P_{\nu_{\alpha}\to\nu_{\beta}}(x,t)
=
\left|
\langle\nu_{\beta}|\nu_{\alpha}(x,t)\rangle
\right|^2
=
\left|
\sum_{k}
U_{{\alpha}k}
\,
e^{- i E_k t + i p_k x}
\,
U_{{\beta}k}^*
\right|^2
\,,
\label{004}
\end{equation}
which is manifestly Lorentz invariant.

Considering ultrarelativistic neutrinos,
we apply now the assumption \ref{A3},
$t = x = L$,
where $L$ is the distance traveled by the neutrino between production and detection.
The phase in Eq.~(\ref{004}) becomes
\begin{equation}
E_k t - p_k x
=
\left( E_k - p_k \right) L
=
\frac{ E_k^2 - p_k^2 }{ E_k + p_k } \, L
=
\frac{ m_k^2 }{ E_k + p_k } \, L
\simeq
\frac{ m_k^2 }{ 2 E } \, L
\,.
\label{005}
\end{equation}
It is important to notice that Eq.~(\ref{005})
shows that the phases of massive neutrinos relevant for the oscillations
are independent from any assumption on the energies and momenta
of different massive neutrinos,
as long as the relativistic dispersion relation
$ E_k^2 = p_k^2 + m_k^2 $
is satisfied.
This is why the standard derivation of the neutrino oscillation probability
gives the correct result,
in spite of the unrealistic equal momentum assumption \ref{A4}.

Using the phase in Eq.~(\ref{005}),
the oscillation probability as a function of the distance $L$
has the standard expression in Eq.~(\ref{104}).
Let us notice that the expression (\ref{104})
is still Lorentz invariant,
as shown in Ref.~\cite{physics/0305122},
because $L$ is not the instantaneous source-detector distance
but the distance traveled by the neutrino between
production and detection.

\section{Wave packet model}
\label{Wave packet model}

Several wave packet models of neutrino oscillations
have been devised, with similar results, in the framework of
Quantum Mechanics
\cite{Giunti-Kim-Lee-Whendo-91,Kiers:1996zj,Kiers-Weiss-PRD57-98,Giunti-Kim-Coherence-98,Giunti:2001kj,Giunti:2003ax}
and Quantum Field Theory
\cite{Giunti-Kim-Lee-Lee-93,Giunti-Kim-Lee-Whendo-98,Cardall-Coherence-99,Beuthe:2002ej,Giunti:2002xg}
(see Ref.~\cite{Beuthe:2001rc} for a comprehensive review).
Here we briefly review the main points of
the quantum field theoretical
wave packet model
presented in Ref.~\cite{Giunti:2002xg},
which is based on the assumption in
Quantum Field Theory
that free particle are described by
wave packets constructed as appropriate
superpositions of states in the momentum Fock space
of the corresponding free field.

The wave packet describing a neutrino
created with flavor $\alpha$
in the process
\begin{equation}
P_I \to P_F + \ell^{+}_{\alpha} + \nu_{\alpha}
\label{601}
\end{equation}
is given by
\begin{equation}
| \nu_{\alpha} \rangle
\propto
\langle P_F , \ell^{+}_{\alpha} |
- i \int \mathrm{d}^4x \,
\mathcal{H}_I(x) \,
| P_I \rangle
\,,
\label{602}
\end{equation}
where we have considered the first order perturbative contribution
of the effective weak interaction hamiltonian
$\mathcal{H}_I(x)$.
The states
$| P_I \rangle$,
$| P_F \rangle$,
$| \ell^{+}_{\alpha} \rangle$
that describe the particles taking part to the localized production process
have the wave packet form
\begin{equation}
| \chi \rangle
=
\int \mathrm{d}^3p
\,
\psi_{\chi}(\vet{p};\vet{p}_{\chi},\sigma_{p\chi})
\,
| \chi(\vet{p},h_{\chi}) \rangle
\qquad
(\chi = P_I , P_F , \ell^{+}_{\alpha})
\,,
\label{604}
\end{equation}
where
$\vet{p}_{\chi}$
is the average momentum,
$\sigma_{p\chi}$
is the momentum uncertainty,
determined by the interactions with the surrounding medium,
and
$h_{\chi}$ is the helicity.
Approximating the momentum distributions
$\psi_{\chi}(\vet{p};\vet{p}_{\chi},\sigma_{p\chi})$
with gaussian functions,
the integrals in the expression (\ref{602})
for the neutrino state
can be calculated analitically,
leading to
\begin{equation}
| \nu_{\alpha} \rangle
=
N_{\alpha}
\sum_{k} U_{{\alpha}k}^{*}
\int
\mathrm{d}^3p
\,
e^{-S^P_{k}(\vet{p})}
\sum_h
\mathcal{A}^P_{k}(\vet{p},h)
\,
| \nu_{k}(\vet{p},h) \rangle
\,,
\label{605}
\end{equation}
where
$N_{\alpha}$
is a normalization factor,
$\mathcal{A}^P_{k}(\vet{p},h)$
is the amplitude of production of a neutrino with mass $m_k$,
momentum $\vet{p}$ and helicity $h$,
and the function
$e^{-S^P_{k}(\vet{p})}$
enforces energy-momentum conservation within
the momentum uncertainty
due to the momentum distributions of
$P_I$,
$P_F$,
$\ell^{+}_{\alpha}$.
The spatial width
$\sigma_{xP}$
of the neutrino wave packet
is related to its momentum uncertainty
\begin{equation}
\sigma_{pP}^2
=
\sigma_{pP_{I}}^2
+
\sigma_{pP_{F}}^2
+
\sigma_{p\ell^{+}_{\alpha}}^2
\label{606}
\end{equation}
by the minimal Heisenberg uncertainty relation
$\sigma_{xP} \sigma_{pP} = 1/2$.

Let us consider the detection of a neutrino with flavor $\beta$ through the
charged-current weak process
\begin{equation}
\nu_{\beta} + D_I \to D_F + \ell_{\beta}^{-}
\,,
\label{607}
\end{equation}
at a space-time distance
$(\vet{L},T)$
from the production process,
where the neutrino 
created with flavor $\alpha$
is described by a state obtained by acting on $| \nu_{\alpha} \rangle$
in Eq.~(\ref{605})
with the space-time translation operator
$\exp( -i \widehat{E} T + i \widehat{\vet{P}} \cdot \vet{L} )$,
where
$\widehat{E}$ and $\widehat{\vet{P}}$
are the energy and momentum operators,
respectively:
\begin{equation}
| \nu_{\alpha}(\vet{L},T) \rangle
=
N_{\alpha}
\sum_{k} U_{{\alpha}k}^{*}
\int
\mathrm{d}^3p
\,
e^{ -i E_{\nu_{k}}(\vet{p}) T + i \vet{p} \cdot \vet{L} }
\,
e^{-S^P_{k}(\vet{p})}
\sum_h
\mathcal{A}^P_{k}(\vet{p},h)
\,
| \nu_{k}(\vet{p},h) \rangle
\,.
\label{608}
\end{equation}
The detection amplitude is given by
\begin{equation}
\mathcal{A}_{\alpha\beta}(\vet{L},T)
=
\langle
D_F , \ell_{\beta}^{-}
|
- i \int \mathrm{d}^4x \,
\mathcal{H}_I(x) \,
|
D_I , \nu_{\alpha}(\vet{L},T)
\rangle
\,.
\label{609}
\end{equation}
The transition probability as a function of
the distance $\vet{L}$ is given by the average over
the unmeasured time $T$
of $|\mathcal{A}_{\alpha\beta}(\vet{L},T)|^2$.
In the realistic case of ultrarelativistic neutrinos
the final result for the flavor transition probability is
\begin{align}
P_{\alpha\beta}(L)
=
\sum_{k} |U_{{\alpha}k}|^2 |U_{{\beta}k}|^2
+
2 \mathrm{Re}
\null & \null
\sum_{k>j} U_{{\alpha}k}^{*} U_{{\beta}k} U_{{\alpha}j} U_{{\beta}j}^{*}
\exp\Bigg[
- 2 \pi i \, \frac{L}{L^{\mathrm{osc}}_{kj}}
\nonumber
\\
\null & \null
- \left( \frac{L}{L^{\mathrm{coh}}_{kj}} \right)^2
- 2 \pi^2 \kappa \left( \frac{\sigma_{x}}{L^{\mathrm{osc}}_{kj}} \right)^2
\Bigg]
\,,
\label{611}
\end{align}
where
$L^{\mathrm{osc}}_{kj} = 4 \pi E / \Delta{m}^2_{kj}$
are the standard oscillation lengths
and
$L^{\mathrm{coh}}_{kj} = 4 \sqrt{2\omega} E^2 \sigma_{x} / |\Delta{m}^2_{kj}|$
are the coherence lengths.
The quantities
$\kappa$
and
$\omega$,
which are usually of order one,
depend on the production and detection processes
\cite{Giunti:2002xg}.
The total spatial coherence width
$\sigma_{x}$
is given by
\begin{equation}
\sigma_{x}^2
=
\sigma_{xP}^2
+
\sigma_{xD}^2
\,.
\label{612}
\end{equation}

The form of the flavor transition probability in Eq.~(\ref{611})
is consistent with the results obtained with other
wave packet models
in the framework of
Quantum Mechanics
\cite{Giunti-Kim-Lee-Whendo-91,Kiers:1996zj,Kiers-Weiss-PRD57-98,Giunti-Kim-Coherence-98,Giunti:2001kj,Giunti:2003ax}
and Quantum Field Theory
\cite{Giunti-Kim-Lee-Lee-93,Giunti-Kim-Lee-Whendo-98,Cardall-Coherence-99,Beuthe:2002ej}.
One can see that the standard value of the oscillation phase
is confirmed, in agreement with the discussion
in the previous Sections.
In addition,
the wave packet treatment produced a coherence term
and a localization term.

The coherence term
$ \displaystyle
\exp[-( L / L^{\mathrm{coh}}_{kj} )^2]
$
suppresses the oscillations due to
$\Delta{m}^2_{kj}$
when
$ L \gtrsim L^{\mathrm{coh}}_{kj} $,
because the wave packets of the massive neutrino components
$\nu_k$ and $\nu_j$
have separated so much that they cannot be absorbed coherently
in the detection process.

The localization term
$ \displaystyle
\exp[- 2 \pi^2 \kappa ( \sigma_{x} / L^{\mathrm{osc}}_{kj} )^2]
$
suppresses the oscillations due to
$\Delta{m}^2_{kj}$
if
$ \sigma_{x} \gtrsim L^{\mathrm{osc}}_{kj} $.
This means that in order to measure the interference
of the massive neutrino components
$\nu_k$ and $\nu_j$
the production and detection processes must be localized
in space-time regions much smaller than the
oscillation length $L^{\mathrm{osc}}_{kj}$.
In practice this requirement is easily satisfied by all
neutrino oscillation experiments,
because the space-time coherence regions
of the production and detection processes
are usually microscopic,
whereas the oscillation length is usually macroscopic.

The localization term is important
for the distinction of
neutrino oscillation experiments
from experiments on the measurement of neutrino masses.
As first shown by Kayser in Ref.~\cite{Kayser:1981ye},
neutrino oscillations are suppressed
in experiments able to measure the value of a neutrino mass,
because the measurement of a neutrino mass
implies that only the corresponding massive neutrino is
produced or detected.

Kayser's \cite{Kayser:1981ye}
argument goes as follows.
Since a neutrino mass
is measured from energy-momentum conservation
in a process in which a neutrino is produced or detected,
from the energy-momentum dispersion relation
$ E_k^2 = p_k^2 + m_k^2 $
the uncertainty of the mass determination is
\begin{equation}
\delta{m_k}^2
=
\sqrt{ \left( 2 E_k \delta{E_k} \right)^2 + \left( 2 p_k \delta{p_k} \right)^2 }
\simeq
2 \sqrt{2} E \sigma_p
\,,
\label{022}
\end{equation}
where the approximation holds for realistic ultrarelativistic neutrinos
and
$\sigma_p = 1 / 2 \sigma_x$
is the momentum uncertainty.
If
$
\delta{m_k}^2
<
|\Delta{m}^2_{kj}|
$,
the mass of $\nu_k$ is measured
with an accuracy better than the difference
$\Delta{m}^2_{kj}$.
In this case
the neutrino $\nu_j$ is not produced or detected
and the interference of
$\nu_k$ and $\nu_j$
is not observed.
The localization term
in the oscillation probability
(\ref{611})
automatically implements Kayser's mechanism,
because
$ \sigma_{x} / L^{\mathrm{osc}}_{kj} $
can be written as
$ \Delta{m}^2_{kj} / 4 \sqrt{2} E \sigma_p $.
If
$
\delta{m_k}^2
<
|\Delta{m}^2_{kj}|
$,
the localization term in Eq.~(\ref{611})
suppresses\footnote{
Using the arguments
presented by M. Beuthe in Ref.~\cite{Beuthe:2002ej},
it is possible to show that
$\kappa$
is always of order one.
}
the interference of
$\nu_k$ and $\nu_j$.

\section{Conclusions}
\label{Conclusions}

In conclusion,
we have shown that
the probability of neutrino oscillations
can be derived in a covariant way
in the plane wave approach
starting from realistic assumptions.
We have also presented a derivation of neutrino oscillations
in a quantum field theoretical wave packet approach.
In both cases we obtained the standard expression for
the oscillation phase.
The wave packet approach allows also
to describe the coherence of the oscillations
and the localization of the production
and detection processes.

\end{document}